\documentclass{aa}       
\usepackage{graphics,graphicx}

\def\spose\rlap
\def\simless{\mathbin{\lower 1pt\hbox
   {$\spose{\raise 5pt\hbox{$\char'074$}}\char'430$}}}
\def\simgreat{\mathbin{\lower 1pt\hbox
   {$\spose{\raise 5pt\hbox{$\char'076$}}\char'430$}}}
\def\simgreat{\gapp}
\def\simless{\lapp}
\def\lapp{\mathbin{\raise2pt \hbox{$<$} \hskip-9pt \lower4pt \hbox{$\sim$}}}
\def\gapp{\mathbin{\raise2pt \hbox{$>$} \hskip-9pt \lower4pt \hbox{$\sim$}}}

\begin{document}
 
   \thesaurus{11.01.2, 11.03.4, 11.09.1, 11.14.1, 13.25.2} 
%
\title{BeppoSAX observation of Hercules A and MRC 0625-536}

  \author{E. Trussoni\inst{1}, L. Feretti\inst{2}, S. Massaglia\inst{3},
P. Parma\inst{2} }
 
   \institute{Osservatorio Astronomico di Torino, Strada Osservatorio 20, 
 I-10025 Pino Torinese, Italy   
\and 
Istituto di Radioastronomia del CNR, Via Gobetti 101, I-40129 Bologna, Italy
\and
Dipartimento di Fisica Generale dell'Universit\`a,
Via Pietro Giuria 1, I-10125 Torino, Italy  }

\offprints{E. Trussoni}
 
   \date{Received; accepted;}
 \titlerunning{BeppoSAX observations of Her A and MRC 0625-536}
 
\authorrunning{Trussoni et al.} 
 
\maketitle
 
\begin{abstract} 
We present BeppoSAX observations of the two FR~I type radio galaxies  
Hercules A  (3C 348) and  MRC 0625-536 in the energy range $0.2 - 200$ keV. 
Data analysis  shows that the X-ray flux from Hercules A is consistent with
a diffuse thermal 
plasma emitting at $T \approx 4-5$ keV with a possible, but somewhat 
uncertain, contribution of a softer component at $T \approx 3 $ keV. The 
non thermal  emission from the active nucleus must be significantly smaller 
than the thermal  one, and no indication of relevant core obscuration by 
a surrounding torus  was detected.
The flux from MRC 0625-536 originates form an extended region and has been 
fitted to a thermal law with $T \approx
5.7$ keV and with a column density consistent with the galactic absorption.
A spatially resolved spectral analysis does not  
show a relevant variation of the temperature and the metallicity across the
diffuse emission zone.  A non thermal spectral component,  
related to the nuclear activity,
may be  present in the innermost region with some 
possible amount of local obscuration, contributing 
 $\lapp \; 10\%$ to the total luminosity. Hard X-ray emission  
from  MRC 0625-536 has been detected in the PDS (15 - 200 keV) that may 
be related  either to its galactic core or to the intracluster region.

\keywords{galaxies: active - galaxies: clusters: individual: A 3391 - 
galaxies: individual: Hercules A; MRC 0625-536 - galaxies: nuclei - 
X-rays: galaxies}
      
\end{abstract}

\section{Introduction}

The two sources Hercules A (3C 348) and MRC 0625-536 are radio galaxies   
classified as Fanaroff-Riley I (FR~I) type objects, although
Hercules A has indeed the morphology of a FR~I radio galaxy, but with a much 
higher  radio luminosity, typical of a FR~II object.  The galaxy associated 
with MRC 0625-536 is the main member of the cluster A 3391, while  Hercules A 
lies in a group.

Recent spectral and morphological analyses of 
 X-ray data, collected by  ROSAT and ASCA (Otani et al. 1998, Siebert et 
al. 1999, Gliozzi et al. 2000), have provided  information on the physical and 
dynamical properties of the environment surrounding these objects and of their 
active nuclei. Both radio galaxies are embedded in diffuse clouds of X-ray 
emitting hot plasma of size larger than the extended, radio emitting, regions.
The active nucleus may also contribute to the X-ray flux, as observed in
several FR~I radio galaxies, and this emission is likely 
related to the inner region of the relativistic jet (Worrall \& 
Birkinshaw 1994, Trussoni et al. 1999, Hardcastle \& Worrall 1999). 

In this paper we present  BeppoSAX observations of Hercules A and  
MRC 0625-536. This analysis provides new information on the structure 
of these radio galaxies and their associated clusters. In fact, the 
instrumentation on board of BeppoSAX allows us to collect data, 
during the same pointing, that range from the soft band 
up to $\approx 200$ keV. We will also examine the variation of the spectral 
parameters of the X-ray emission across the clusters,
relying upon the smoother (with respect to the ASCA detectors)
radial behavior of the Point Spread Function of the BeppoSAX MECS detectors.

In Section 2 we summarize the known properties of the two sources, 
in Section 3 we outline the details of the observations
and of the data analysis while in Section 4 we report 
the results of the spectroscopic analysis. Their implications
on the structure of the two sources are discussed in detail
in Section 5,  taking into account also previous observations, and 
in Section 6 we summarize our results.
 A value of $H_0= 50$ km s$^{-1}$ Mpc$^{-1}$ is 
assumed throughout.
 
\section{The sources}

We list in Table 1 the main properties of Hercules A and MRC 0625-536
and recall in the following the  available literature.

\begin{table*}
{\caption{Main data on the targets}}
\begin{tabular}{|c|c|c|c|c|c|c|c|c|c|c|c|c|}
\hline
Source & RA (J2000) & DEC (J2000) & $z$ &  $m_{\rm V}$  & $P^{\rm a}_{\rm tot}$
& $P^{\rm a}_{\rm core}$ & Galaxy & Environment \\
\hline
Hercules A  & 16$^h$ 51$^m$ 08.1$^s$ & 04$^{\circ}$ 59$^{\prime}$
34$^{\prime\prime}$ &  0.154  &  16.9  & 150 & 0.12 & Giant cD  & Group  \\
MRC 0625-536 &  06$^h$ 26$^m$ 15.4$^s$ & -53$^{\circ}$ 40$^{\prime}$
52$^{\prime\prime}$   & 0.054  &  15.4  & 2.50  &  0.056 & Pair/Dumbbell
&   A 3391   \\
\hline
\end{tabular}
 
$^{\rm a}$ $\times 10^{25}$  W Hz$^{-1}$ at 5 GHz

\label{tab:table}
\end{table*}
 
\begin{table*}
{\caption{Observational details}}
\begin{tabular}{|c|c|c|c|c|c|c|c|}
\hline
Source  & Obs. date & t$_{{\rm exp}, {\rm MECS}}$  &   $t_{{\rm exp}, {\rm
LECS}}$ & t$_{{\rm exp}, {\rm PDS}}$  & Source cts$^{\rm a, b}_{\rm MECS}$
&  Source cts$^{\rm b}_{\rm LECS}$ & count rate$^{\rm b}_{\rm PDS}$ \\
\hline
Hercules A     & 28/3/1997 &  18790 s   & 9875 s  &  8550 s
               & $ 1477 \pm 48 $   &  $ 432 \pm 30 $
               &   -  \\
MRC 0625-536   & 8/11/1996  & 13386 s  & 3588 s   &  5136 s
               & $4213 \pm 70$   &  $ 368 \pm 29$
               &  $0.43 \pm 0.13$  \\
\hline
\end{tabular}
           
$^{\rm a}$  merging the counts of the three instruments
 
$^{\rm b}$  Source - background $\pm 1 \sigma$
 
\label{tab:table}
\end{table*}
                  
\noindent
\subsection{Hercules A (3C 348)}

This source is a giant cD elliptical (Spinrad et al. 1985, Zirbel 1997) 
embedded in an environment that  Zirbel  (1997) has classified as
a group of the first Bautz-Morgan class, with Hercules A as the main member. 
At radio frequencies, 3C 348 shows two large and symmetric lobes connected 
by twin jets and without the presence of any hot spot  (Dreher \& 
Feigelson 1984). The radio emission originates in the 
extended lobes mainly,
with a weak contribution of a compact core  (S$_{\rm c,\,4.8 \,
GHz}$ = 10 mJy, corresponding to  $ < 0.1\%$ of the total emission,  
Morganti et al. 1993). Polarization maps show a strong asymmetry in the
fractional  polarization, with the most prominent jet lying in the more 
strongly polarized lobe (Harvanek \& Hardcastle 1998, Gizani \& Leahy 1999). 
This fact can be interpreted as due to the Laing-Garrington effect (Laing 1988,
Garrington et al. 1988) and is consistent with the existence of a cluster/group
environment around the source.  

The X-ray properties of 3C 348, derived by ROSAT and ASCA observations, are 
discussed by Siebert et al. (1999). The flux detected by ASCA  is 
fit  to a thermal model
 with temperature $T \approx 4.6$ keV and metallicity 
$\mu \approx 0.38$ (with  respect to the standard cosmic values). The presence 
of an additional non thermal component, with a flat slope (photon index 
$\Gamma \sim 1.7$), is also consistent with the data. The ROSAT/PSPC 
data  are consistent with a plasma emitting at temperature 
$T \sim 2.4$ keV, suggesting that a two temperature gas may be  
present around Hercules A, without any significant temperature gradient  
up to an outer  radius of 3$^{\prime}$. The total luminosity in 
the energy band 0.1 - 2.4 keV turns out to be $4.4 \times 10^{44}$ 
erg s$^{-1}$.
  
The brightness profile obtained by the  ROSAT/HRI data analysis is consistent 
with a $\beta$-model with $r_c \approx 35^{\prime \prime}$ ($\approx 120$ kpc)
 and $\beta  \approx 0.63$, with the diffuse emission extending up to $\approx 
3^{\prime}$ from the center. A point-like structure, 
coincident with the nucleus of the galaxy, is detected as well
and contributes for a fraction of 
$\approx 8\%$ to the total emission. The luminosity in the soft X-ray band 
(0.1 - 2.4 keV) results in  $3.4 \times 10^{43}$ erg s$^{-1}$ for the nucleus 
(assuming a  power law  spectrum with $\Gamma = 1.7$).

\subsection{ MRC 0625-536} 

This radio source is associated with  the eastern component 
of a pair of galaxies (also classified as a dumbbell galaxy) and is the 
brightest member at the center of the cluster A 3391. The radio morphology 
 shows a wide-angle tail structure (Morganti et al. 1999, and references
therein), with a compact core of flux S$_{\rm c,\,4.8 \,
 GHz}$ = 42 mJy, which
contributes only $\approx 2\%$ to the total emission.

X-ray studies are presented by Otani et al. (1998) and by Gliozzi et al. 
(1999). The ASCA and ROSAT/PSPC data are consistent with a thermal spectrum 
with $T \sim 6 - 6.5$ keV and $\mu \approx 0.35$.  The PSPC observations 
 also show that
the flux from the innermost $2^{\prime}$ is consistent with a power law 
spectrum.

The region of X-ray emission is  extended 
(radius of $\approx 15^{\prime}$) and, from ROSAT/HRI data, its 
brightness profile is fit to
a $\beta$-model with core radius $\approx 2.5^{\prime}$ ($\approx 220$ 
kpc) and $\beta \approx 0.56$, typical of a cluster.
The central nucleus can contribute to the flux 
only up to $\approx 3\%$, with a luminosity (0.1 - 2.4 keV) of $\approx 4.1 
\times 10^{42}$ erg s$^{-1}$ (power law with $\Gamma = 1.67$), while the
luminosity of the extended region is $\approx 1.5 \times 10^{44}$ erg s$^{-1}$.
 
\section{The observations and data analysis} 

The BeppoSAX observation plan, covering the energy range 0.2 - 200
keV, is reported in Table 2 (for details on the BeppoSAX instrumentation see
Boella et al. 1997). The event files of the three MECS and of the LECS have 
been processed with 
the packages FTOOLS 4.2 and SAXDAS 2.0.2. The adopted matrices for the 
instrument response and the effective areas were released in September 97. The 
spectral analysis has been performed with the
XSPEC package (ver. 9.0) rebinning the counts in order to
have at least 20 events in each energy channel.
For the thermal fits the MEKAL model for an optically 
thin plasma has been employed. We have adopted the HI measurements, reported 
in Dickey \& Lockman (1990), for the values of the galactic column density 
($N_{\rm H,gal}$) used in the spectral fits.
 The errors for the parameters are at 68\% of
uncertainty (1$\sigma$). The background, extracted from  
the public BeppoSAX 
archive,  consists  of different merged exposures of blank fields with 
the same extent as the source. We have limited the spectral fits to the 
energy ranges 0.2 - 4  keV  for the LECS, and 1.5 - 9.5 keV for the MECS.

The analysis of extended sources requires the correction of the Point Spread 
Function (PSF) for the vignetting effects. This correction can be 
performed 
for the MECS observations through the command {\it effarea} of the SAXDAS 
package that provides the suitable auxiliary response matrices (Molendi 1998, 
D'Acri et al. 1998). This command requires an `a priori' estimate 
of the radial  distribution of the counts: we have generated the 
corrected matrices 
using the brightness profile derived from the HRI observations (see Section
2). The photons have been extracted, for both sources, from a region with
radius of $8^{\prime}$: this choice is satisfactory for Hercules A, while it 
includes only the central region of the more extended MRC 0625-536. In order to
analyze the spectral properties of different regions of the diffuse emission,
we have also carried out spectral fits to photons extracted from the inner 
region (radius $2^{\prime}$) and external concentric annuli, with size 
dependent  on the count statistics. 
 
At present, no analogous procedure exists for processing the data of LECS 
observations of extended sources. To reduce the vignetting effects in this 
case, we have  extracted
the photons from a smaller region (radius of $6^{\prime}$), 
even though the PSF of the LECS is wider than that of the MECS. Considering 
this limitation, we have not carried out simultaneous fits of the spectra 
of the two instruments. The  LECS data at softer energies have 
therefore been considered as a  consistency 
test of the MECS results and for comparison with the previous ROSAT 
observations.
 
The background subtracted spectra of the PDS have been extracted from 
the public BeppoSAX archive. For the positive detection of MRC 0625-536, 
and in  order to have a significant statistics for a useful analysis, 
the counts have been rebinned to 4 energy channel bins in
the energy range 15 - 200 keV.

\section{Results}

\subsection{Hercules A} 
The results of the spectral analysis are given in Table 3.
The spectrum obtained from the whole emitting region and detected in the MECS, 
with  the column density fixed  to its galactic value ($N_{\rm H,gal}=6.3 
\times  10^{20}$ 
cm$^{-2}$), is consistent with thermal emission (see Fig. 1). The fit 
yields a temperature $T \approx 4.2 - 5.4$ keV and metallicity 
$\mu \approx 0.2 - 0.5$. 
The resulting unabsorbed flux (2 -  10 keV) is  $f = 3.5
 \times 10^{-12}$ erg cm$^{-2}$ s$^{-1}$, corresponding to a 
luminosity of $4.3 \times 10^{44}$ erg s$^{-1}$, and of $4.7 
\times 10^{44}$ erg s$^{-1}$ in the ROSAT energy band (0.1 - 2.4 keV). 

We have then divided the whole region in  concentric annuli
extending up to  $8^{\prime}$ (Table 3).
We have obtained from the thermal fits $T \approx 4 - 5$ keV and $\mu 
\approx 0.5-0.6$. These values are all consistent, within 1$\sigma$ 
fluctuations, with emission from a homogeneous and isothermal
gas (see Table 3).
In the external zone $4^{\prime}-8^{\prime}$, we were only able
to derive the flux,
after setting $T=5$ keV and $\mu=0.4$, since only
$\approx 15\% $ of the total counts are emitted in this outer region.
We thus conclude that the present data do not show any significant 
variation of the spectral parameters across the intragroup plasma. 

Some contribution to the flux emitted from the central region
(of $2^{\prime}$ radius, $\approx 700$ photons) originates from the 
galactic nucleus. This emission is quite low: from the 
parameters derived for the power law emission  from  ASCA and ROSAT 
($\Gamma = 1.7$,  $f = 3.3 \times 10^{-13}$ erg cm$^{-2}$ s$^{-1}$) 
we expect in the MECS a count rate about 
ten times lower than the thermal component. Consistently a single 
power law spectrum does not fit with the  emission from this region 
($\chi^2 \approx 1.5$). Performing a fit with a composite model 
(thermal + power law) by  fixing
$T=4$ keV, $\mu=0.4$ and $\Gamma = 1.7$ we found for the non thermal emission
an upper limit $\approx  2.5$ times larger than the luminosity
detected in Siebert et al. (1999). This  confirms
that no strong X-ray variability occurred in the nucleus of Hercules A between 
the observations of  August 1993 (ROSAT/PSPC), August/September 1996 
(ROSAT/HRI), March 1997 (BeppoSAX) and August 1998 (ASCA).

\begin{table*}
\caption{Spectral fits for Hercules A$^{\rm a}$}
\label{tab:table}
\begin{tabular}{|c|c|c|c|c|c|c|}
\hline
Region &   $T$ (keV)    & $\mu$ &  $A_{\rm th}$ (cm$^{-5}$)  &  $\chi^2$ 
(d.o.f.) \\ 
\hline
 $0^{\prime}$ - $8^{\prime}$   & $4.78^{+0.67}_{-0.55}$ &                        
                   $0.36^{+0.16}_{-0.14}$ & $5.49^{+0.44}_{-0.40} \times         
                    10^{-3}$  & 0.94 (70)\\
\hline
 $0^{\prime}$ - $2^{\prime}$   & $4.31^{+0.57}_{-0.48}$ &                        
                    $0.64^{+0.24}_{-0.20}$ &         
                      $3.90^{+0.37}_{-0.33} \times 10^{-3}$  & 1.07 (28) \\
                    & $4.55^{+0.56}_{-0.46}$ &  0.4 &  $4.08^{+0.35}_{-0.32}    
                       \times 10^{-3}$  & 1.09 (29) \\
\hline
$2^{\prime}$ - $4^{\prime}$   & $5.20^{+1.12}_{-0.83}$ &  $0.55^{+0.31}_{-0.26}$ 
                    &  $9.52^{+0.98}_{-1.04} \times 10^{-4}$  & 0.89 (23) \\
                    & $5.39^{+1.09}_{-0.79}$ &  0.4   &  $9.68^{+0.96}_{-0.86}    
                   \times 10^{-4}$  & 0.87 (24) \\    
\hline
$4^{\prime}$ - $8^{\prime}$   & 5 &   0.4                    
                  & $3.88 \pm 0.67 \times 10^{-4}$  & 0.92 (29) \\  
\hline
\end{tabular} 
 
$^{\rm a}$ $N_{\rm H} = N_{\rm H,gal}$; quantities without errors are fixed  


\end{table*}

\begin{figure}[t]
\rotatebox{270}{\resizebox{2.5in}{!}{\includegraphics{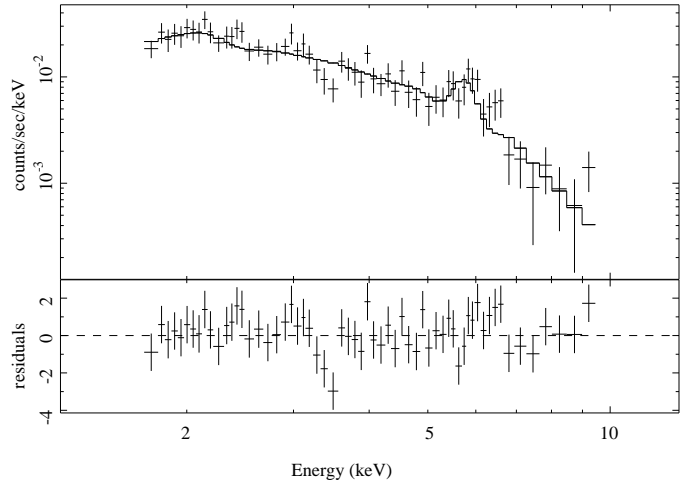}}}
\vspace{0.5cm}
\caption{Thermal spectrum (MECS) of the whole region of Hercules A (radius
$8^{\prime}$) assuming $N_{\rm H}=N_{\rm H,gal}$.  }
\end{figure}

The flux in the LECS was fitted to a thermal spectrum without a satisfactory 
estimate of the parameters,  the column density only could be constrained
yielding $N_{\rm H}= 6.75^{+3.3}_{-2.3} \times 10^{20}$ cm$^{-2}$ 
($\chi^2 = 1.13$, 15 d.o.f.),  in agreement with the galactic hydrogen 
column density.
Thus setting  $N_{\rm H} =  N_{\rm H,gal}$ and $\mu=0.4$ we derived 
$A_{th} = 4.60^{+0.31}_{-0.32}$ cm$^{-5}$ and a 
temperature $T = 3.12^{+0.70}_{-0.49}$ keV, with an upper limit (at $2 \sigma$ 
level) $T \leq 4.45$ keV ($\chi^2 = 1.01$). Apparently the temperature in this 
softer band results lower. 

The lack of any detected flux in the PDS is in agreement with the extrapolation
of the thermal (from the cluster) and non thermal (from the core) spectra
to the 15 - 200 keV energy band. We expect in the energy range of the PDS 
a flux from the cluster of $\sim
10^{-13}$ erg cm$^{-2}$ s$^{-1}$ and from the non thermal component
(allowing some amount of local absorption, as discussed below) $\approx
2 \times 10^{-12}$ erg cm$^{-2}$ s$^{-1}$. Both these fluxes are well below
the detection threshold of the PDS.

The value of the temperature obtained in the MECS 
is in excellent agreement with the ASCA observation, while the
lower value deduced from the LECS data seems to confirm the PSPC
results. Concerning the non thermal nuclear 
emission, the comparison of the flux detected in the HRI with
the upper limit found in the MECS implies also an upper limit for the
local column density: $ \simless \; 4 \times 10^{21}$ cm$^{-2}$ 
($90 \%$ of confidence).
Then the obscuration of the core by an edge-on projected disk is not relevant.
This is also consistent with the very high luminosity
detected by the HRI in the soft band, higher by a factor $\approx 10$
than expected from the correlation between the radio and X-ray core
luminosities in FR~I radio galaxies (and $\approx 3$ times larger than
its $90 \%$ upper confidence limit; Canosa et al. 1999, Hardcastle \& Worrall
1999). 

We will discuss these points further in Section 5.1.

\begin{table*}
\caption{Spectral fits for MRC 0625-536$^{\rm a}$}
\label{tab:table}
\begin{tabular}{|c|c|c|c|c|c|c|}
\hline
Region &   $T$ (keV)    & $\mu$ &  $A_{\rm th}$ (cm$^{-5}$)  &  $\chi^2$ 
(d.o.f.) 
&  $\Gamma$   & $A_{\rm pl}^{\rm b}$ \\ 
\hline
$0^{\prime}$ - $8^{\prime}$   & $5.66^{+0.40}_{-0.35}$ &                        
                   $0.33^{+0.08}_{-0.07}$ & $1.98^{+0.07}_{-0.06} \times         
                    10^{-2}$  & 1.02 (108) &  & \\
\hline
$0^{\prime}$ - $2^{\prime}$   & $7.60^{+1.54}_{-1.13}$ & $0.13^{+0.16}_{-0.13}$ & 
                     $4.81^{+0.31}_{-0.31} \times 10^{-3}$  & 0.75 (34) &  & \\
                    &  $7.10^{+1.32}_{-0.97}$ &  0.33 &  $4.64^{+0.30}_{-0.28}    
                       \times 10^{-3}$  & 0.75 (35) &  & \\
\hline
$2^{\prime}$ - $4^{\prime}$   & $5.50^{+0.68}_{-0.56}$ &  $0.47^{+0.15}_{-0.14}$ 
                &  $6.11^{+0.37}_{-0.36} \times 10^{-3}$  & 0.95 (50) &  & \\
                    & $5.72^{+0.67}_{-0.55}$ &  0.33 &  $6.24^{+0.35}_{-0.32}    
                     \times 10^{-3}$  & 0.95 (51) &  & \\
\hline
$4^{\prime}$ - $6^{\prime}$  & $4.79^{+0.69}_{-0.55}$ &  $0.49^{+0.19}_{-0.17}$ 
                &  $4.96^{+0.37}_{-0.34} \times 10^{-3}$  & 1.03 (42) &  & \\    
                    & $5.05^{+0.67}_{-0.55}$ &  0.33 &  $5.06^{+0.36}_{-0.33}    
                     \times 10^{-3}$  & 1.03 (43) &   & \\
\hline
$6^{\prime}$ - $8^{\prime}$ & $6.75^{+1.88}_{-1.19}$ &  $0.42^{+0.26}_{-0.24}$ 
                &  $3.29^{+0.29}_{-0.28} \times 10^{-3}$  & 0.69 (32) &  & \\   
                    & $6.97^{+1.81}_{-1.17}$ &  0.33 &  $3.32^{+0.27}_{-0.24}    
                     \times 10^{-3}$  & 1.03 (43) &   & \\
\hline     
$0^{\prime}$ - $2^{\prime}$ & 5.7 &  0.33             
                           & $2.93^{+1.08}_{-1.43} \times 10^{-3}$                    
                           & 0.72 (35) &  1.7 & $5.00^{+3.35}_{-2.53} \times     
                           10^{-4}$\\ 
\hline     
$0^{\prime}$ - $2^{\prime}$ & 5.7 &  0.33             
                           & $4.18^{+0.58}_{-0.52} \times 10^{-3}$                    
                           & 0.74 (35) &  1.2 & $1.10^{+0.54}_{-0.73} \times     
                           10^{-4}$\\ 
\hline
\end{tabular} 

$^{\rm a}$ $N_{\rm H} = N_{\rm H,gal}$; quantities without errors are fixed  
 
$^{\rm b}$ photons cm$^{-2}$ s$^{-1}$ keV$^{-1}$  (at 1 keV)
\end{table*}

\begin{figure}[t]
\rotatebox{270}{\resizebox{2.5in}{!}{\includegraphics{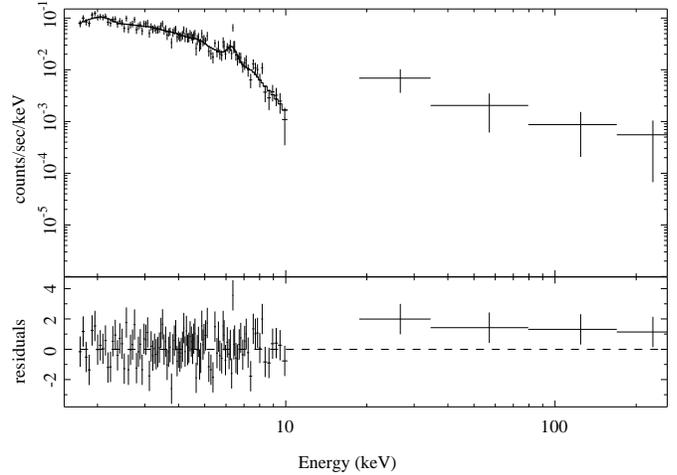}}}
\vspace{0.5cm}
\caption{Thermal spectrum (MECS + PDS) of MRC 0625-536 assuming $N_{\rm 
H}=N_{\rm H,gal}$.  
The fit has been performed only to the MECS data.   
}
\end{figure}
 
\subsection{MRC 0625-536}

The flux detected in the MECS in the whole region ($8^{\prime}$ in radius) is 
fit to a thermal emission model with  $T \approx 5.7$ keV and $\mu \approx 
0.3$, assuming $N_{\rm H} \equiv N_{\rm H,gal} = 5.4 \times 10^{20}$ cm$^{-2}$ 
(see Fig. 2). The total 
unabsorbed flux is  $f=1.7 \times 10^{-11}$ erg cm$^{-2}$ s$^{-1}$, with 
luminosity of $2.2 \times 10^{44}$ erg cm$^{-2}$ s$^{-1}$, in both the  2 - 
10 keV and 0.1 - 2.4 keV bands. 

After selecting four concentric circular regions with spacing $\Delta r = 
2^{\prime}$ we carried out fits to the emitted fluxes deriving
values of the temperature and
metallicity basically consistent with the `average' values  obtained analyzing 
the whole cluster region (see Table 4 and Fig. 3). We obtain a similar 
value for the temperature $T$ 
setting  $\mu=0.33$ in the fit. In the innermost region  we can notice 
an increase in temperature and a decrease in metallicity, that may be 
a signature of non thermal emission from  the active nucleus.

To test this point further, we have verified that the central flux  also fit  
 a power law model with $\Gamma = 1.84^{+0.08}_{-0.08}$
($\chi^2=0.77$). The total number of photons from this zone ($\approx 900$) 
does  not allow a complete analysis with a composite spectrum.  
Setting $T=5.7$ keV, 
$\mu =0.33$ and $\Gamma=1.7$ we obtain acceptable fits ($\chi^2=0.72$) 
with fluxes of the two components
(2 - 10 keV, with large errors; see Table 4) $f_{\rm th} \approx 2.5 
\times 10^{-12}$ erg cm$^{-2}$ s$^{-1}$ and $f_{\rm pl} \approx 2.0 \times 
10^{-12}$ erg cm$^{-2}$ s$^{-1}$, respectively. This  corresponds to 
a  nuclear  luminosity of $L_{\rm pl} \approx 2.6  \times 10^{43}$ 
erg s$^{-1}$ (a similar value of $L_{\rm pl}$ is obtained in the softer band 
0.1 - 2.4 keV).

Little useful information can be obtained from the LECS  observation: setting
$\mu=0.33$ we derive  $T = 5.09^{+5.60}_{-1.76}$ keV and 
$N_{\rm H} = 1.71^{+0.95}_{-0.74} \times 10^{21}$ cm$^{-2}$ 
($\chi^2 = 1.09$, 12 d.o.f.). This value is larger than the galactic column 
density by a factor $\sim 2 - 3$, however its lower limit (at 2$\sigma$ 
level), $5.5 \times 10^{20}$ cm$^{-2}$, is coincident with $N_{\rm H,gal}$. 

The flux  detected in the high energy band by the PDS (at $\approx 3 
\sigma$ level) is well above the extrapolated
thermal emission  from the cluster and from the active
nucleus, assuming the parameters derived from the MECS observation. In fact, 
the fluxes expected  in the 15 - 200 keV  energy range are $\approx 1.1 
\times 10^{-12}$ erg cm$^{-2}$ s$^{-1}$ and $\approx 2.0 \times 10^{-12}$ erg 
cm$^{-2}$ s$^{-1}$, respectively, below the detection capability  
of the PDS (see Fig. 2).  

\begin{figure}[t]
\resizebox{\hsize}{!}{\includegraphics{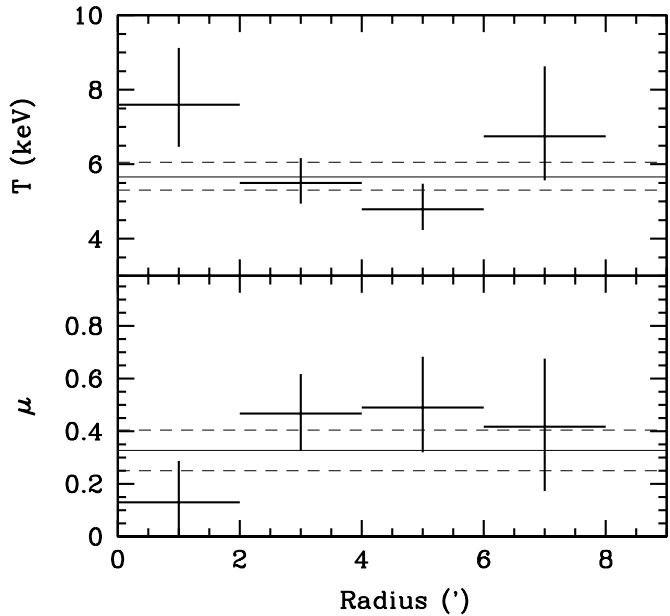}}
\vspace{0.5cm}
\caption{Temperature (upper panel) and metallicity  (lower panel)  
vs the radial  distance in MRC 0625-536. The thin horizontal lines 
indicate the values of
$T$ and $\mu$ (solid) $\pm 1 \sigma$ (dotted) deduced from the fit to  
the whole region ($r = 8^\prime$).}
\end{figure}

The values of the temperature and metallicity obtained by the BeppoSAX 
observations are fully consistent with those of ASCA and PSPC (Otani et al. 
1998) within  $1 \sigma$ statistical fluctuations. The total luminosity from 
the cluster in the 2 - 10 keV range is $\approx 90\%$ lower than derived from 
ASCA consistently with the slightly smaller region of photon extraction 
assumed for the  MECS. 

The  nuclear luminosity derived from the HRI observation   
(Gliozzi et al. 1999) exceeds by a factor $\approx 3$  the one predicted by
the correlation of Canosa et al. (1999), but is still within 90\% 
of uncertainty.
This would suggest the absence of heavy local absorption 
 in the nucleus of this radio galaxy.
However, the MECS data indicate a much higher luminosity 
(by a factor $\approx$ 6) with respect
to HRI, but, due to the large statistical fluctuations, 
the lower limit of the luminosity
in the MECS is $\sim$ three times the one in 
the HRI. Assuming no variability, this discrepancy could be ascribed
to a local column density of $\sim 5 \times 10^{21}$ cm$^{-2}$
surrounding the core. This point will  be discussed further in the next 
Section, considering also the results of the PDS observation.

\section {Discussion}

We outline the main implications of the BeppoSAX observations considering
also previous observations from other satellites.

\subsection {Hercules A}

Our results confirm the
discrepant value of the temperature of the thermal diffuse emission obtained
from the ASCA and ROSAT observations and 
support the possibility that a two temperature plasma is present in Hercules
A. However,  since both MECS and PSPC data are consistent with emission 
from an isothermal medium,  it is difficult to gain some insights about
the origin of these two thermal 
components. Unfortunately  it is not possible to merge 
the LECS and MECS data to perform a spectral fit with a two temperature model
(see Section 3).

Some  clues for the interpretation of this phenomenology  may be provided 
by the HRI image (Siebert et al. 1999), that shows peculiar morphologies, 
unrelated to the extended thermal gas and suggested to originate from the 
interaction of the expanding radio lobes with the hot medium. We can then 
argue that this processed gas has physical
conditions far from those of a plasma in ionization 
equilibrium.  On the other hand,  the extent of  this irregular structure 
is smaller but comparable with the size of the diffuse 
region modeled with a $\beta$ profile ($\sim 2^{\prime}$, see Fig. 4 of 
Siebert et al. 1999 ), such that  the imaging capability of the MECS 
does not allow to spatially disentangle these different regions.
   
It is worth noticing that the luminosity and temperature of the thermal
plasma detected in Hercules A are consistent  
with the correlation found for the galaxy clusters  (Wu et al. 1999).
Therefore, the X-ray properties of this object are typical
of rich clusters, despite the fact that  Hercules A has been associated 
with a group of galaxies, which would be expected to show
a lower temperature and  luminosity (Zirbel 1997, Ponman et al. 1996).

The MECS data confirm also that only a small fraction of emission from the 
central region can originate from the core, that however  
it is very bright when compared with similar objects. 
It turns out that in fact that the X-ray luminosity of the nucleus of 
Hercules A is typical of that of FR~II narrow lines radio galaxies ($4 
\times 10^{17}$ W Hz$^{-1}$ sr$^{-1}$, see Fig. 6 in  Hardcastle 
\& Worrall 1999). This may mean 
that extra  contributions to the core emission could come from other 
components,  e.g. from an thick accretion  disk (we cannot exclude also 
that the inner  nucleus is actually obscured by
the torus and the non thermal X-ray emission originate from the jet on
much larger scales). On the other hand, the optical and radio 
core luminosities are consistent with the properties of FR~I
radio galaxies where obscuration does not play a relevant role
and no thick accretion disks appear to be present (Chiaberge et al. 1999). 
However the above argument implies that in this  
radio  galaxy the low value of the core radio luminosity, with respect to the 
total one, may not be a reliable indicator of local absorption in the 
nucleus.  

In conclusion, Hercules A appears very  peculiar at X-ray 
energies on large and small scales: observations from the new generation 
telescopes (Chandra) will provide new insights on this interesting object.

\subsection{MRC 0625-536}
 
The MECS data confirm the spectral properties  obtained by the previous 
observations with ASCA and ROSAT.   The values of the 
temperature and metallicity are typical for a cluster, and the luminosity
is again consistent with the  $L_{\rm X}$-$T$  correlation (Wu et al. 1999). 
The lack of any variation of the temperature across the region of emission 
confirms that this cluster has very likely undergone a merging process (see 
also Otani et al. 1998). 

The detection of hard X-ray flux in the PDS, even though weak,
may have important implications on the interpretation of this radio galaxy.  
X-ray emission above 20 keV is commonly observed in Seyfert 
galaxies (see e.g. Matt 1998).
Among clusters, hard  X-ray emission has been  found so 
far only in Coma and A 2256 with good significance, and in A 2199 with
marginal significance (Fusco-Femiano et al. 1999, 2000, Kaastra et al.  1999)
 
As stated in Sect. 4.2, the high energy emission cannot be the extrapolation
of the thermal emission from the intracluster plasma, detected 
from the MECS. Should this emission originate from a hotter thermal 
component, the fit to the PDS data would require a temperature  
$T \geq 80$ keV. We can reasonably exclude the presence 
in this cluster of a plasma with these extreme physical conditions.

For investigating the origin of the hard emission, we have first looked for   
possible contamination by a different source in the PDS field of view.
The nearby cluster A 3395 falls within the PDS field of view,
 but its thermal emission  could not increase the flux by more than a 
factor $\sim 2$ (Henriksen \& Jones 1996). 
A BL-Lac object, MS 0622.5-5256, is present in the PDS field, at the position
RA$_{2000}$ = 06$^h$ 23$^m$ 39.1$^s$, DEC$_{2000}$ = -52$^{\circ}$ 
57$^{\prime}$ 49$^{\prime\prime}$, i.e. at about 49$^{\prime}$ from the field 
center. The flux of this source in the 0.3 - 3.5 keV energy band is $3.7 \times
10^{-13}$ erg cm$^{-2}$ s$^{-1}$, estimated assuming a
power-law spectrum with a photon index $\Gamma$=2 (Stocke et al. 1991).
Extrapolating this flux to the PDS energy band, and correcting for the 
attenuation due to the distance of the BL-Lac from the field center,
it turns out that it is about 100 times lower than the flux detected in
the PDS and therefore cannot be responsible for it, unless an unusually
strong variability in the flux of the BL-Lac object has occurred. 

Thus, it seems reasonable to consider that the PDS flux is of non-thermal 
origin, and actually originating from the target itself. We consider in 
the following  three main possibilities : 1) it is related to the nucleus 
of the radio galaxy MRC 0625-536; 2) it is related to the radio lobes of 
MRC 0625-536; 3) it is associated to the cluster A 3391. 

1) In this case, the hard X-ray emission would originate from the core of 
radio galaxy MRC 0625-536. A similar case is represented by the radio galaxy
NGC 1275 (3C 84), in the Perseus cluster, 
where hard non thermal emission has been  detected by OSO7 (Rothschild et al. 
1981).  BeppoSAX observations have confirmed the existence of 
X-ray emission at  energies exceeding 20 keV, with flux of $\sim 4 \times 
10^{-11}$ erg cm$^{-2}$ s$^{-1}$ in the 20 - 100 keV band  (Molendi 1998). 
According to the MECS analysis and to the results of previous 
observations, the emission from the core of MRC 0625-536
is not detectable in the PDS energy range. One might, however, speculate
on the existence of a component with a much flatter spectral index:
for $\Gamma = 1.2$ one would derive a (15 - 200 keV) flux 
of $ 7.2 \pm 4.6 \times 10^{-11}$ erg cm$^{-2}$ s$^{-1}$, corresponding to a 
luminosity of $\approx 9.0 \times 10^{44}$ erg s$^{-1}$.  
Fitting the MECS flux from the central region (radius $2^{\prime}$) 
to a composite spectrum with the  same parameters as in Table 3 for the thermal
component, and with $\Gamma = 1.2$ for the power law spectrum, one obtains 
for the latter a flux $1.3^{+0.7}_{-0.9} \times 10^{-11}$ erg cm$^{-2}$ 
s$^{-1}$ in the 15 - 200 keV band, $\approx 5 $ times less than detected in 
the PDS. However, considering the large statistical fluctuations, the 
difference  between the upper limit of the MECS flux and the lower limit 
of the PDS flux (90\% of  confidence) reduces to a factor $\approx 1.3$. 
A larger discrepancy is found with the HRI data on the nuclear emission, that 
is $\approx $  3 times less than the lower limit of the PDS flux. 
 This difference disappears if some amount of obscuration  is present
with  $N_{\rm H} \approx 5 - 6  \times 10^{21}$  cm$^{-2}$, as deduced 
in Section 4.2.  In this case  the intrinsic core 
luminosity in the soft band would increase by a factor $\sim 2$. 

This interpretation has two major difficulties. First, 
the nucleus would result, in the X-ray band,  overluminous with respect 
to its radio power (see Sec. 4.2). The second  problem is the extreme 
flatness of the  spectrum. In blazars,
where the hard X-ray emission is related to the inverse Compton process
in the relativistic jet, it is generally found that $\Gamma \gapp 1.5$ for 
luminosities $\sim 10^{44 - 45}$ erg s$^{-1}$ (and steeper spectra in the 
$\gamma$ energy band; see Fossati et al. 1998). It has 
 been alternatively proposed
 that advection dominated disks (ADAF) could contribute 
to the X-ray core emission of low power radio galaxies (Sambruna et al. 1999), 
with a very flat spectral emission (see e.g. Allen et al. 2000), but 
with a luminosity much lower  (a factor $ < 0.01$) than appropriate for 
our target. 

2) The  hard X-ray emission might be produced by Inverse Compton
scattering of relativistic electrons   within the radio lobes
with the photons of the Cosmic Microwave Background (CMB),  as detected
in a few radio galaxies (e.g., Feigelson et al. 1995, Tsakiris et al. 1996).
A non-thermal power law model with $\Gamma$ = 2, gives for the present
source a flux $\sim$ 4 $\times$ 10$^{-11}$ erg cm$^{-2}$ s$^{-1}$ in the
15-200 keV band. Using  the total radio flux
at 4.8 GHz of  1.85 Jy (Morganti et al. 1993) and the radio spectral
index  $\sim$1 (Otani et al. 1998), the
detected X-ray flux would imply a magnetic field of about 0.25 $\mu$G,
i.e. more than an order of magnitude lower than the equipartition value
derived from the data published by Otani et al. (1998).
With this low magnetic field, the relativistic electrons emitting in
the radio domain up to 8.87 GHz
(frequency to which the radio spectrum extends,
Otani et al. 1998) would have very high energy, largely sufficient to
account for the 15-200 keV Inverse Compton emission.
Although there
is no firm evidence in the literature that radio sources should be
in equipartition conditions, there is also no evidence of such
large deviations from it.
 Therefore, this interpretation seems unlikely.      

3) Non thermal X-ray emission  from the 
cluster A~3391 would be expected if relativistic
electrons present in the intergalactic medium undergo
Inverse Compton scattering  with the
photons of the CMB. This kind of emission has
been  detected by BeppoSAX in the two clusters Coma (Fusco-Femiano 
et al. 1999) and A 2256 (Fusco-Femiano et al. 2000), where 
the presence of ultra-relativistic electrons in the intracluster medium
is directly demonstrated by the existence of diffuse radio halos.
Conversely, no diffuse radio emission is
detected in the cluster A 3391, so the presence of ultra-relativistic 
electrons in the intergalactic medium remains questionable. 
In any event, if the hard X-ray emission is of Inverse Compton origin, 
the magnetic field must be very weak in order to avoid producing 
an observable radio halo. A similar conclusion is reached for the
cluster A 2199, where a marginal detection of hard X-rays above the expected
thermal emission is reported by Kaastra et al. (1999), and no 
diffuse radio halo is observed. For the cluster A 2199, 
Kempner \& Sarazin (1999)
suggest the alternative possibility that the X-ray emission 
above 20 keV could be due to non thermal bremsstrahlung by a population of 
suprathermal electrons. The same interpretation might apply to the present 
case.

We conclude that, even though reasonable, none of the possibilities 
that we have analyzed for the interpretation of the PDS emission is 
fully convincing. More  data would be needed to clarify this issue. 
 
\section{Summary}

We summarize the main results obtained from the BeppoSAX observations and
their implications on the properties of these two radio galaxies and their
environment. 

\noindent
{\it Hercules  A.} No relevant variation of temperature and metallicity occurs
between the inner and outer regions of the cluster. However a plasma with
two different temperatures seems to be present, perhaps related
to the interaction of the radio components with intracluster gas. It must be 
borne in mind anyway that the structure of this object  is quite
peculiar: at optical wavelengths it is  classified as a group but its
X-ray properties are typical of clusters. 

The non thermal emission from the active nucleus appears unabsorbed, as 
expected for this class of radio galaxies, but brighter than expected
by its core radio emission. This could be related to the peculiar
nature of Hercules A that shows properties of both FR~I and FR~II
radio galaxies.

\noindent
{\it MRC 0625-36.} The X-ray luminosity and the values of $T$ and $\mu$ 
are consistent with those expected for this kind of clusters. The lack
of strong variation of the temperature and metallicity  across the emitting 
region suggests that no cooling flow is present, while it could be related 
to a recent merging process. The nuclear non thermal emission detected in 
the MECS and in previous  missions is consistent with that expected from 
radio data,  without excluding however some amount of local obscuration. 

Concerning finally the high energy flux detected in the PDS, it seems 
unlikely that it is related to the cluster or to the radio lobes. On the 
other hand, a nuclear origin for this emission would imply a 
luminosity of some units of $10^{44}$ erg s$^{-1}$ in the hard band.
Either this emission is completely independent of the flux detected at 
lower energies, or, to fit the ROSAT, ASCA and BeppoSAX (MECS) data, it 
 must have an very flat spectrum.

\begin{acknowledgements}

The authors thank M. Capalbi, F. Fiore and S. Molendi for their 
suggestions in the treatment of the MECS observations of extended sources, 
and  G. Brunetti, R. Fanti, M. Hardcastle, R. Morganti, P. Padovani and F. Vagnetti for 
discussions and comments on the manuscript. We  
warmly acknowledge the late Daniele Dal Fiume, 
for his invaluable  help in the treatment of the  PDS data. This work was 
partially supported by the Italian Ministry for University and
Research (MURST), under grant Cofin98-02-32,  and by the Italian Space
Agency (ASI).

\end{acknowledgements}


\begin{thebibliography}{}

\bibitem{}Allen S.W., Di Matteo T., Fabian A.C., 2000, MNRAS 311, 493 
 
\bibitem{}Boella G., Butler R.C., Perola G.C., Piro L., Scarsi L.,
Bleeker J.A.M.,  1997, A\&AS 122, 299
 
\bibitem{}Canosa C.M., Worrall D.M.,
 Hardcastle M.J., Birkinshaw M.,
1999, MNRAS 310, 30
 
\bibitem{}Chiaberge M., Capetti A., Celotti A., 1999, A\&A 349, 77
 
\bibitem{}D'Acri F., de Grandi S., Molendi S., in:
The Active X-ray Sky: Results from BeppoSAX and RXTE.
L. Scarsi, H. Bradt, P. Giommi, and F. Fiore eds;
Nuclear Physics B (Proc. Suppl.), vol. 69,  p. 581
 
\bibitem{}Dickey J.M., Lockman F.J., 1990, ARA\&A 28, 215

\bibitem{}Dreher J.W., Feigelson E.D., 1984, Nat 308, 43

\bibitem{}Feigelson E.D.,
 Laurent-Muehleisen S.A.,
 Kollgaard R.I., Fomalont E.B.,
1995, ApJ 449, L149

\bibitem{}Fossati G., Maraschi L., Celotti A., Comastri A.,
Ghisellini G.,
1998, MNRAS 299, 433

\bibitem{}Fusco-Femiano R.,
 dal Fiume D., Feretti L.,
 Giovannini G., Grandi P.,
 Matt G., Molendi S.,
 Santangelo A.,
1999, ApJ 513, L21
 
\bibitem{}Fusco-Femiano R.,
 Dal Fiume D.,
 De Grandi S., Feretti L.,
 Giovannini G., Grandi P.,
 Malizia A., Matt G.,
 Molendi S.,
2000, ApJ 534, L7
 
\bibitem{}Garrington S.T., Leahy J.P., Conway R.G.,
Laing R.A., 1988, Nat 331, 147

\bibitem{}Gizani N.A.B., Leahy J.P., 
1999, New Astr. Rev. 43, 639

\bibitem{}Gliozzi M., Brinkmann W.,
Laurent-Muehleisen S.A.,
Takalo L.O., Sillanp\"a\"a A.,
1999, A\&A 352, 437
 
\bibitem{}Hardcastle M.J., Worrall D.M.,
1999, MNRAS 309, 969
                     
\bibitem{}Harvanek M., Hardcastle M.J., 
1998, ApJS 119, 25

\bibitem{}Henriksen M., Jones C.,
1996, ApJ 465, 666
 
\bibitem{}Kaastra J.S., Lieu R.,
 Mittaz J.P.D., Bleeker J.A.M., Mewe R.,
 Colafrancesco S., Lockman F.J.,
1999, ApJ 519, L119
 
\bibitem{}Kempner J.C., Sarazin C.L.,
1999, American Astron. Soc. Meeting 195, p. 1014

\bibitem{}Laing R.A., 
1988, Nat 331, 149

\bibitem{}Matt G., 1998, in:
The Active X-ray Sky: Results from BeppoSAX and RXTE.
L. Scarsi, H. Bradt, P. Giommi, and F. Fiore eds;
Nuclear Physics B (Proc. Suppl.), vol. 69,  p. 467
 
\bibitem{}Molendi S., 1998, in:
The Active X-ray Sky: Results from BeppoSAX and RXTE.
L. Scarsi, H. Bradt, P. Giommi, and F. Fiore eds;
Nuclear Physics B (Proc. Suppl.), vol. 69,  p. 563

\bibitem{}Morganti R., Killeen N.E.B., Tadhunter C.N., 1993, MNRAS 263, 1023

\bibitem{}Morganti R., Oosterloo T., Tadhunter C.N., Aiudi R., Jones P., 
Villar-Martin M.,
1999, A\&AS 140, 355 

\bibitem{}Otani C., Brinkmann W.,
 Boehringer H., Reid A., Siebert J.,
1998, A\&A 339, 693
 
\bibitem{}Ponman T.J., Bourner P.D.J., Ebeling H., B\"horinger H., 1996,
MNRAS 283, 690
 
\bibitem{}Rothschild R.E., Baity W.A.,
 Marscher A.P., Wheaton W.A.,
1981, ApJ 243, L9                                                               
\bibitem{}Sambruna R.M., Eracleus M., Mushotzky R.F.,
1999, ApJ 526, 60

\bibitem{}Siebert J., Kawai N., Brinkmann W.,
1999, A\&A 350, 25
 
\bibitem{}Spinrad H., Marr J., Aguilar L.,
 Djorgovski S.,
1985, PASP 97, 932
 
\bibitem{}Stocke J.T., Morris S.L., Gioia I.M., et al., 1991,
ApJS 76, 813


\bibitem{}Trussoni E., Vagnetti F.,
 Massaglia S., Feretti L., Parma P.,
 Morganti R., Fanti R., Padovani P.,
1999, A\&A 348, 437
 
\bibitem{}Tsakiris D., Leahy J.P., Strom R.G., Barber C.R.,
1996, in: Extragalactic Radio Sources, R. Ekers, C. Fanti, 
and L. Padrielli eds; Kluwer Academic Publisher, p. 256

\bibitem{}Worrall D.M., Birkinshaw  M., 1994, ApJ 427, 134                       
\bibitem{}Wu X.-P., Xue Y.-J.,
 Fang L.-Z.,
1999, ApJ 524, 22
 
\bibitem{}Zirbel E.L.,
1997, ApJ 476, 489
 
\end{thebibliography}
\end{document}